\documentclass[preprint,12pt]{elsarticle}

\usepackage{graphicx}
\usepackage{float}
\usepackage[utf8]{inputenc}
\usepackage{color}
\usepackage{appendix}
\usepackage{multirow}
\usepackage[unicode]{hyperref} 

\hypersetup{colorlinks,urlcolor=black,citecolor=black,linkcolor=black,filecolor=black}

\usepackage{subcaption}
\usepackage{enumitem}
\usepackage{amssymb}
\usepackage{amsmath}

\usepackage{soul}
\usepackage[normalem]{ulem}
\usepackage{orcidlink}
\newcommand{\orcidGB}{\orcidlink{0000-0003-2849-0120}} 
\newcommand{\orcidGGB}{\orcidlink{0000-0001-9223-6480}} 
\newcommand{\orcidGP}{\orcidlink{0000-0001-5038-678X}} 
 
\newcommand{\orcidZJ}{\orcidlink{0009-0001-2300-3605}} 

\journal{Medical Image Analysis}

\begin{document}

\begin{frontmatter}

\title{Proton Computed Tomography Image Reconstruction Based on the Richardson\,--\,Lucy Algorithm}

\author{Gábor Bíró$^{1}$\orcidGB}

\author{Ákos Sudár$^{3, 4}$}

\author{Zsófia Jólesz$^{1,2}$\orcidZJ}

\author{Gábor Papp$^2$\orcidGP}

\author{Gergely Gábor Barnaföldi$^{1}$\orcidGGB}

\author{for the Bergen pCT collaboration}

\affiliation{
organization={$^{1}$HUN-REN Wigner Research Centre for Physics},
addressline={29--33 Konkoly--Thege Mikl\'os \'ut},
city={Budapest},
postcode={H-1121},
country={Hungary},
}
\affiliation{
organization={$^{2}$ELTE E\"otv\"os Lor\'and University, Institute of Physics},
addressline={1/A P\'azm\'any P\'eter S\'et\'any},
city={Budapest},
postcode={H-1117},
country={Hungary},
}
\affiliation{
organization={$^{3}$National Institute of Oncology},
addressline={Ráth György utca 7-9},
city={Budapest},
postcode={H-1122},
country={Hungary},
}
\affiliation{
organization={$^{4}$Budapest University of Technology and Economics, Institute of Nuclear Techniques},
addressline={Műegyetem rakpart 9},
city={Budapest},
postcode={H-1111},
country={Hungary},
}

\begin{abstract}

Proton therapy is an emerging method in cancer therapy. One of the main developments is to increase the accuracy of the Bragg-peak position calculation, which requires more precise relative stopping power (RSP) measurements. A promising choice is the application of proton computed tomography (pCT) systems which takes the images under similar conditions, as they use the same irradiation device and hadron beam for imaging and treatment. A key aim is to develop a precise image reconstruction algorithm for pCT systems to reach their maximal performance.

In this work, an iterative image reconstruction algorithm, based on the Richardson\,--\,Lucy iteration is proposed for the first time for proton CT image reconstruction.
Monte Carlo (MC) simulations of CTP528 and CTP404 phantoms were used to benchmark the  proposed method.
In the case of an idealized detector setup, using a 1mm pitch grid, 4.88~lp/cm spatial resolution and 0.66\% average RSP uncertainty was achieved. 
The present method provides a promising proof-of-concept candidate for compromise between accuracy and speed with several further development directions.
\end{abstract}




\end{frontmatter}

\section{Introduction} 

Hadron therapy is an emerging and efficient treatment method against cancer. The increasing number of hadron therapy centers and the number of successful treatments demonstrate its success since the first proposal in the 50s~\citep{protonBerkely}. Today's accelerator techniques led us to use protons or even light ions as bombarding particles. The application of massive hadron beams instead of massless X-ray results in more focused dose distribution~\citep{Durante2017}. Indeed, using higher mass number ion beams than protons (He, C and O) can result in increased relative biological effectiveness (RBE) in the tumor volume~\citep{Durante2021}.
The higher the dose gradient around the treated volume is required, the lower the uncertainty 
in the relative stopping power (RSP) distribution during dose planning is needed, to avoid insufficient dosage of the tumor or the overdose of organs at risk~\citep{VESTERGAARD2023100441, paganetti_uncert, 10.1002/mp.15644, 10.1088/0031-9155/60/19/7585}.

\begin{figure}[h]
    \centering
    \includegraphics[width=0.48\textwidth]{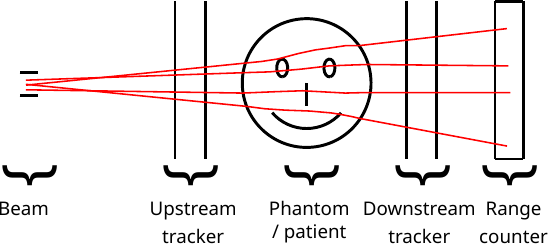}
    \caption{Detector design of the list mode imaging concept.}
    \label{fig:concepts}
\end{figure}

The developments of proton Computed Tomography (pCT) techniques are promising solutions for the above problems. Applying the same irradiation device, beam, and hadron for both the medical imaging and the treatments can significantly reduce the uncertainties of the imaging.
To obtain this, two main imaging strategies exist:
\begin{enumerate}[label=(\roman*)]
    \item The first concept is to measure the average energy loss of the proton beam. This design is feasible from a technical point of view, but the achievable spatial resolution is poor with the clinically available proton beams \citep{extendedMLPformalism}. 
    \item The second concept is the so-called {\it list mode} imaging concept, which measures the energy loss and in parallel it estimates the path of each individual proton. Monte Carlo (MC) simulations and prototype measurements showed that this solution can meet the required spatial and density resolutions, so the focus is moved toward this direction~\citep{pCTReview2018}.
\end{enumerate}
Nowadays, the pCT scanner R\&Ds around the world are tending to reach the prototyping and clinical/pre-clinical testing phase, which requires the integration of the prototype scanners into clinical environment~\citep{BergenpCTStatusReport}.
Following the list mode strategy, the path estimation of individual protons is usually based on the measurements of the upstream and downstream tracker detector pairs, which concept is called the {\it double-sided scanner design}, as illustrated on Figure~\ref{fig:concepts}~\citep{9427964, ProtonVDA}.
One important further step can be the abandonment of the upstream tracker detectors, and the application of a more affordable {\it single-sided scanner design}. The drawback of this latter concept is the less accurate proton path measurement, however the study by~\cite{noFrontTrack_Rambo} concluded that the achievable spatial resolution meets the minimum requirement. Anyhow, the spatial resolution remains a point to improve of this solution, so even a small resolution increase would be important for the application in the clinical practice. The more precise modeling of the measurement inaccuracy in the system matrix is a possible way to reach this goal.

The realistic clinical applicability requires to complete the data taking and processing within minutes, which results in 1-10 million protons per second measurement rate. The ultimate goal would be to finish the image capturing within the minimal gantry rotation time. The LLU/UCSC Phase-II Scanner prototype detector showed data taking speed up to 1.2 million proton per second, which probably can be upgraded by 50\% in the near future~\citep{LL-Phase2}. In order to increase the data taking rate to 10 million protons per second, two possible directions exist: the first is to apply faster readout frequency of 10 MHz at least, while the second is to measure multiple proton tracks within one readout frame. The second solution fits mostly with the single-sided scanner design, because this solution avoids the pairing problem of the upstream and downstream measurements, which leads to track confusion in case of a double-sided scanner even with low number of protons in a frame. 

Multiple proton measurements fit best for silicon pixel trackers and silicon pixel sensor based range counters as presented by the Bergen pCT Collaboration~\citep{high-granularityDTC, DesignPixelRangeTelescope, BergenpCTStatusReport, AliceAlpide, mapspaper}. However, multiple proton measurements can also be done by applying three silicon strip detectors rotated relative to each other as presented by the PRaVDA collaboration~\citep{PRaVDA}. Another layout has been designed by the iMPACT group~\citep{8335784}. The ProXY detector combines the two acceleration possibilities with monolithic active pixel detectors~\citep{mapspaper}. Applying this layout, about 50 MHz readout frequency can be reached, and it is planned to measure multiple proton events~\citep{ProXY}.

In this paper a novel imaging method is proposed for these pCT detector concepts. The  Richardson\,--\,Lucy algorithm~\citep{Richardson,LBLucy} originates from astronomy and was found to be the maximum-likelihood expectation-maximization (ML-EM) solution of the image reconstruction of emission tomography~\citep{Shepp-Vardi-1982}. Richardson\,--\,Lucy-based deconvolution methods already have been used to improve the quality of degraded medical images for X-ray CT, particularly those affected by blur and noise~\citep{Yaqoub_Al-Ani_2023}. However, in this study an algorithm based on the the Richardson\,--\,Lucy method is used for proton CT imaging for the first time. In this paper we demonstrate, that the proposed method can be a promising candidate for future medical applications.

The paper is organized  as follows: section~\ref{sec:methodology} begins with the general approach to the image reconstruction problem itself, followed by the brief presentation of the details of the Richardson\,--\,Lucy algorithm and the proton-phantom interaction model. Section~\ref{sec:test} compares detector designs and presents our proposed image reconstruction algorithm. Section~\ref{sec:test} also contains the evaluation of the spatial- and density resolution of phantoms. Results are summarized and discussed in sections~\ref{sec:results} and~\ref{sec:discussion}, respectively.

\section{Image Reconstruction Techniques} 
\label{sec:methodology}

The primary objective of image reconstruction in proton computed tomography (pCT) is to determine the three-dimensional distribution of the relative stopping power (RSP) from measured proton data, most notably the water-equivalent path length (WEPL). Owing to multiple Coulomb scattering and energy-loss straggling, proton trajectories deviate significantly from straight lines, rendering classical reconstruction approaches developed for X-ray CT non-trivial to apply. Consequently, pCT reconstruction methods are commonly grouped into two broad classes: analytical (filtered backprojection–based) methods and iterative (algebraic) methods.

\subsection{Analytical Reconstruction Methods}

Analytical reconstruction methods are conceptually derived from filtered backprojection (FBP), where the image is reconstructed by integrating projection data along assumed particle paths. In the context of pCT, the simplest straight-line approximation is inadequate because it neglects multiple Coulomb scattering at beam energies $\mathcal{O}(100\ \textrm{MeV})$. To address this limitation, modified backprojection schemes have been proposed, most notably the distance-driven backprojection method, which incorporates the curvature of the proton most-likely path (MLP) into the reconstruction process~\citep{pCTRadon}. 

While such approaches can achieve relatively good spatial resolution, they require very high proton statistics to suppress noise and artifacts arising from scattering-induced uncertainties. This requirement translates directly into increased imaging dose, which limits their clinical applicability. As a result, purely analytical reconstruction methods are generally regarded as suboptimal for high-accuracy pCT imaging, particularly when dose efficiency is a critical consideration.

\subsection{Iterative (Algebraic) Reconstruction Methods}

Iterative reconstruction techniques model the imaging process as the interaction between individual proton trajectories and volumetric image elements (voxels). This leads naturally to a large, sparse system of linear equations,
\begin{equation}
    \mathbf{y} = \mathbf{A}\,\mathbf{x},
    \label{eq:lineq}
\end{equation}
where $\mathbf{y}$ is a vector of measured WEPL values for $m$ proton tracks, $\mathbf{x}$ contains the unknown RSP values for $n$ voxels, and $\mathbf{A}$ is the system matrix whose elements are the path lengths of the protons through the voxels, representing the interaction coefficients between tracks and voxels. 
In realistic pCT scenarios, $m \sim 10^{8}$--$10^{9}$ and $n \sim 10^{5}$--$10^{7}$, making the system heavily overdetermined and computationally demanding.

Algebraic Reconstruction Technique (ART) and its variants, including Parallel ART (PART), Simultaneous ART (SART), and diagonally relaxed orthogonal projection (DROP) methods, have been widely applied in pCT~\citep{pCTReview2018, ANDERSEN198481, DROPmethod, fundCompTomo, TVS_DROP_Penfold, Plautz2016}. These methods can be interpreted as iterative row-action or block-iterative solvers that aim to minimize a quadratic loss function of the form
\begin{equation}
    \chi^2 = \frac{1}{2}\sum_i \left( y_i - \sum_j A_{ij} x_j \right)^2,
\end{equation}
often with additional relaxation or regularization terms. Their main strengths lie in their ability to incorporate curved proton trajectories, typically via MLP or spline-based path estimates, and to achieve acceptable spatial and density resolution with moderate proton statistics.

Other studies have demonstrated ART-based pCT reconstructions in both two and three dimensions~\citep{BOUDJELAL2017385, 8347020}. Early work focused predominantly on two-dimensional reconstructions with relatively coarse resolution, often using simplified detector models and limited angular sampling. More recent studies extended ART to 
carbon-ion tomography using simulated data and OpenMP-parallelized CPU implementations~\citep{AVENIDO2022126}. While such approaches demonstrate feasibility and reasonable reconstruction quality, their computational cost remains substantial, particularly when MLP modeling is included explicitly.

\subsection{Acceleration Strategies and GPU Implementations}

To mitigate the computational burden of algebraic methods, various acceleration strategies have been proposed. Ordered-subsets expectation–max\-i\-miza\-tion (OS-EM)~\citep{HudsonLarkin1994} and block-Kaczmarz methods~\citep{Eggermont1981} partition the data into subsets or blocks, significantly accelerating convergence while largely preserving reconstruction quality. These ideas have been successfully exploited in emission tomography and have analogues in transmission imaging.

Hardware acceleration, particularly using graphical processing units (GPU), has proven especially effective. For X-ray CT, GPU-based SART implementations have demonstrated reconstruction times of only a few seconds for phantoms with a resolution of $512 \times 512$ and $1024 \times 1024$ pixels~\citep{8347020}. In contrast, CPU-based ART implementations for pCT typically require minutes to hours for comparable problem sizes, especially in three dimensions. Fully three-dimensional pCT reconstructions incorporating realistic scattering models remain computationally challenging, and GPU-based implementations for such scenarios are still relatively scarce in the literature~\citep{Collins-Fekete_2016}.

\subsection{Motivation for a Richardson--Lucy–Based Approach}

The Richardson\,--\,Lucy (RL) deconvolution algorithm~\citep{Richardson, LBLucy}, originally developed in optics and later widely adopted in emission tomography~\citep{Shepp-Vardi-1982}, is an iterative maximum-likelihood method derived under the assumption of Poisson-distributed data.
It is formulated as a fixed-point iteration that naturally enforces non-negativity of the reconstructed image and exhibits excellent parallelization properties.

Despite its success in emission imaging, to our knowledge, RL has not been previously applied to proton computed tomography. This is notable, given that RL-type algorithms can be interpreted as expectation–maximization schemes and share conceptual similarities with OS-EM methods, which are known to accelerate convergence by processing data in subsets. Furthermore, RL iterations are particularly well suited to GPU architectures, as they rely primarily on sparse matrix–vector operations and element-wise updates.

In this work, we propose and investigate a pCT image reconstruction method based on a generalized Richardson\,--\,Lucy iteration, in which the conventional point-spread function is replaced by the pCT system matrix $\mathbf{A}$. This approach leads to a fundamentally different update rule compared to ART-based methods, while retaining the ability to incorporate curved proton trajectories and realistic physics modeling. The present study is intended as a feasibility and proof-of-concept investigation, with particular emphasis on algorithmic formulation, convergence properties, and suitability for high-performance computing architectures. Similarly to ART studies, we start with a simplified, 2-dimensional case, which serves as a proof-of-concept for the future 3-dimensional extension. A detailed description of the adapted Richardson--Lucy algorithm and its mathematical properties is given below.

\subsubsection{Classical Richardson--Lucy Iteration}

The RL algorithm was originally derived for image deconvolution problems in optics and astronomy~\citep{Richardson, LBLucy}. It is commonly interpreted as a maximum-likelihood estimator under the assumption of Poisson-distributed measurements. In its classical form, the measured intensity $d_i$ at detector pixel $i$ is modeled as
\begin{equation}
    d_i = \sum_j p_{ij} \, u_j ,
\end{equation}
where $u_j$ denotes the unknown source intensity at pixel $j$, and $p_{ij}$ is the point-spread function (PSF), describing the probability that a photon emitted from pixel $j$ is detected at pixel $i$. The PSF satisfies
\begin{equation}
    p_{ij} \ge 0, \qquad \sum_i p_{ij} = 1 ,
\end{equation}
ensuring flux conservation. Assuming Poisson statistics for $d_i$, the RL iteration follows as
\begin{equation}
    u_j^{(k+1)} = u_j^{(k)} \sum_i p_{ij}
    \frac{d_i}{\sum_l p_{il} u_l^{(k)}} .
    \label{eq:RL_classical}
\end{equation}
This iteration is a fixed-point scheme that preserves non-negativity and monotonically increases the Poisson log-likelihood. 

\subsubsection{Linear Forward Model in Proton CT}

In proton computed tomography, the forward problem differs fundamentally from convolution-based imaging. The measured quantity is the water-equivalent path length (WEPL) of individual proton tracks. The measurement model can be written as
\begin{equation}
    y_i = \sum_j A_{ij} \, x_j ,
    \label{eq:pct_forward}
\end{equation}
where $y_i$ is the WEPL of proton track $i$, $x_j$ is the relative stopping power (RSP) of voxel $j$, and $A_{ij}$ quantifies the interaction between track $i$ and voxel $j$. In practice, $A_{ij}$ is non-negative and sparse, and represent (up to a given accuracy) the path length of track $i$ inside voxel $j$ or a weighted contribution accounting for trajectory uncertainty~\citep{Penfold2010ImageRA}.

Unlike the classical RL setting, the index sets of $y_i$ (tracks) and $x_j$ (voxels) are distinct, and the system matrix $\mathbf{A}$ is not shift-invariant and does not represent a convolution. Nevertheless, Eq.~(\ref{eq:pct_forward}) has the same algebraic structure as the forward model used in emission tomography, which motivates adapting the RL framework to pCT.

\subsubsection{Adapted Richardson--Lucy Iteration}

Following the structure of Eq.~(\ref{eq:RL_classical}), we define the predicted WEPL for track $i$ at iteration $k$ as
\begin{equation}
    h_i^{(k)} = \sum_j A_{ij} \, x_j^{(k)} .
    \label{eq:hi_def}
\end{equation}
The adapted Richardson\,--\,Lucy update for the voxel values then reads
\begin{equation}
    x_j^{(k+1)} =
    x_j^{(k)} \,
    \frac{1}{\sum_i A_{ij}}
    \sum_i A_{ij} \,
    \frac{y_i}{h_i^{(k)}} ,
    \label{eq:RL_pct}
\end{equation}
for all voxels $j = 1,\dots,n$.

This form is obtained by identifying the system matrix $\mathbf{A}$ with a generalized, non-shift-invariant transfer operator and introducing an explicit normalization factor $\sum_i A_{ij}$. The normalization ensures dimensional consistency and plays a role analogous to the flux-conservation condition in the classical RL algorithm. If desired, one may equivalently work with a normalized matrix
\begin{equation}
    \tilde{A}_{ij} = \frac{A_{ij}}{\sum_i A_{ij}} ,
\end{equation}
in which case Eq.~(\ref{eq:RL_pct}) reduces formally to the standard RL update with $\sum_i \tilde{A}_{ij} = 1$.

Equation~(\ref{eq:RL_pct}) defines a fixed-point iteration: if $x_j^{(k+1)} = x_j^{(k)}$ for all $j$, then $h_i^{(k)} = y_i$ and the forward model Eq.~(\ref{eq:pct_forward}) is exactly satisfied, independently of the statistical properties of WEPL. 

\subsubsection{Cost Function and Fixed-Point Properties}

Although the classical RL algorithm is derived from a Poisson log-likeli\-hood, in the present transmission-imaging context the statistical interpretation is less direct. Nevertheless, Eq.~(\ref{eq:RL_pct}) can be shown to correspond to a fixed-point solution of a weighted least-squares–type functional. A convenient choice is
\begin{equation}
    \chi^2(\mathbf{x}) =
    \sum_i \frac{\left( y_i - h_i \right)^2}{h_i},
    \qquad
    h_i = \sum_j A_{ij} x_j .
    \label{eq:chi2_def}
\end{equation}
Taking the gradient with respect to $x_k$ yields
\begin{equation}
    \frac{\partial \chi^2}{\partial x_k}
    = -2 \sum_i \frac{y_i - h_i}{h_i} A_{ik}
    + \mathcal{O}\!\left( \frac{(y_i - h_i)^2}{h_i^2} \right) .
    \label{eq:chi2_grad}
\end{equation}
Neglecting higher-order terms close to the solution, a gradient-descent step with step size $\alpha_k$ gives
\begin{equation}
    x_k^{(k+1)} =
    x_k^{(k)} +
    2 \alpha_k
    \sum_i A_{ik}
    \left( \frac{y_i}{h_i^{(k)}} - 1 \right) .
\end{equation}
Choosing a voxel-dependent step size
\begin{equation}
    \alpha_k = \frac{1}{2}
    \frac{x_k^{(k)}}{\sum_i A_{ik}},
    \label{eq:alpha_choice}
\end{equation}
leads exactly to the multiplicative update rule in Eq.~(\ref{eq:RL_pct}). While this choice breaks the strict interpretation as a classical gradient-descent method due to its direction-dependent step size, it preserves positivity and yields a descent direction for $\chi^2$ in the vicinity of the fixed point.

The Hessian of $\chi^2$ near the solution is given by
\begin{equation}
    \frac{\partial^2 \chi^2}{\partial x_k \partial x_l}
    =
    2 \sum_i
    \frac{y_i}{h_i^2}
    A_{ik} A_{il} ,
\end{equation}
which is positive semi-definite for $A_{ij} \ge 0$ and $y_i > 0$. This ensures local stability of the fixed point and convergence under sufficiently small perturbations.

\subsubsection{Mathematical Properties and Practical Implications}

The adapted Richardson--Lucy iteration in Eq.~(\ref{eq:RL_pct}) exhibits several important properties:
\begin{enumerate}[label=(\roman*)]
    \item \textbf{Non-negativity:} If $x_j^{(0)} \ge 0$ for all $j$, then $x_j^{(k)} \ge 0$ for all subsequent iterations.
    \item \textbf{Fixed-point convergence:} The iteration converges to a solution of Eq. (\ref{eq:lineq}) under mild regularity conditions.
    \item \textbf{Noise amplification:} In the classical RL, late iterations tend to amplify statistical noise, necessitating early stopping or regularization~\citep{rl_noise}. In this work we also adopted such techniques in order to optimize the process.
    \item \textbf{Parallelizability:} The update consists of sparse matrix–vector products and element-wise operations, making it well suited for GPU and many-core architectures.
\end{enumerate}

These properties motivate the use of the adapted Richardson\,--\,Lucy algorithm as a viable alternative to ART-based reconstruction in proton CT, particularly in high-performance computing environments. In the following sections, we investigate the numerical behavior, convergence characteristics, and reconstruction quality of the proposed method using simulated pCT data.

\subsection{Sources of Uncertainty in Proton CT Reconstruction}
\label{subsec:uncertainty}

Uncertainties in proton computed tomography arise from multiple, conceptually distinct sources that propagate through the reconstruction pipeline and ultimately limit image quality. For clarity, these sources can be grouped into three broad categories: 

\begin{enumerate}[label=(\roman*)]
    \item data acquisition–related uncertainties, 
    \item physical interaction processes, and 
    \item reconstruction- and model-related uncertainties. 
\end{enumerate}    
While the present work focuses on algorithmic aspects of image reconstruction, understanding these uncertainty sources is essential for interpreting reconstruction performance and for defining meaningful stopping and regularization strategies.

\subsubsection{Data Acquisition and Detector-Related Uncertainties}

The first category (i) comprises uncertainties originating from the detector system and readout electronics. These include electronic noise, finite spatial and energy resolution of tracking and range detectors, digitization effects, and calibration inaccuracies. Together, these effects limit the achievable precision of WEPL measurements and directly propagate into RSP noise in the reconstructed image.

For a fixed detector design, many of these uncertainties can be systematically mitigated through careful calibration, alignment procedures, and signal processing techniques~\citep{extendedMLPformalism, pCTReview2018, noFrontTrack_Rambo}. In simulation-based studies, such as the present work, detector-related effects can be selectively enabled or disabled, allowing the investigation of algorithmic performance under idealized conditions. This separation is particularly useful for isolating reconstruction-induced artifacts from measurement-induced noise.

\subsubsection{Physics-Driven Uncertainties}

The dominant source of uncertainty in pCT arises from the stochastic nature of proton interactions with matter (ii). Multiple Coulomb scattering leads to significant deviations from straight-line trajectories, while energy-loss straggling and nuclear interactions introduce non-Gaussian fluctuations in WEPL. These effects fundamentally limit spatial resolution and may introduce structured artifacts if not properly accounted for~\citep{paganetti_uncert, secondary_fragments, Dickmann_2019}.

The proton most-likely path (MLP) formalisms provide an effective means of mitigating the impact of multiple Coulomb scattering by estimating the expected curved trajectory of each proton based on measured entry and exit conditions~\citep{extendedMLPformalism, MLPChargedParticle, MLP_Schneider, MLP_Schulte}. The accuracy of the MLP model directly affects the fidelity of the system matrix $\mathbf{A}$ and therefore the reconstructed RSP distribution.

Nuclear interactions produce secondary fragments that typically undergo large-angle scattering and exhibit anomalous energy loss. These events violate the assumptions of continuous energy deposition and are a major source of outliers in WEPL measurements. A common and effective mitigation strategy is statistical filtering, such as $3\sigma$ cuts on WEPL residuals, which removes protons likely affected by inelastic nuclear interactions at the cost of reduced statistics~\citep{secondary_fragments, MLP_Schulte}. This trade-off between data purity and statistical noise is intrinsic to pCT imaging.

\subsubsection{Reconstruction and Model-Dependent Uncertainties}

The third category encompasses uncertainties introduced by modeling choices and reconstruction parameters. These include the discretization of the image volume into voxels, angular and spatial sampling of the proton beam, assumptions underlying the system matrix $\mathbf{A}$ (e.g.\ path-length vs.\ weighted interactions), and the treatment of curved trajectories~\citep{TVS_DROP_Penfold, Collins-Fekete_2016, Penfold2010ImageRA, MLPChargedParticle, MLP_Schneider, MLP_Schulte}. Additionally, algorithmic choices such as iteration stopping criteria, subset ordering, and implicit or explicit regularization significantly influence reconstruction quality~\citep{HudsonLarkin1994, Erdogan_1999, penfold_techniques}.

Iterative algorithms, including ART- and Richardson\,--\,Lucy-based methods, are known to exhibit noise amplification in later iterations~\citep{rl_noise, OLIVEIRA2016189}.
Consequently, early stopping based on suitable metrics (e.g.\ WEPL residuals, image variance, or comparison to ground truth in simulation studies) is essential to balance bias and variance. Regularization strategies, whether heuristic or formally derived, can further stabilize the solution but may introduce additional bias.

\subsubsection{Impact on Image Quality Metrics}

All aforementioned uncertainty sources affect the principal image quality metrics in pCT: RSP accuracy (bias), RSP precision (noise), spatial resolution, and the presence of structured artifacts. While advanced techniques (including machine learning-based approaches) offer promising avenues for uncertainty mitigation, well-established physics-based and algorithmic methods already provide effective control over these effects~\citep{penfold_techniques, bdudasml, 10.1088/2632-2153/ae352b}.

The goal of the present study is not to exhaustively optimize all uncertainty sources, but rather to demonstrate the feasibility and mathematical consistency of a novel reconstruction approach based on the Richardson\,--\,Lucy iteration. A systematic investigation of uncertainty mitigation strategies within this framework is deferred to future, dedicated studies.

\subsection{The Proton-Phantom Interaction}
\label{subsec:protonPath}

Instead of the most simple straight line approximation, the estimated (most-likely) path of the protons is used in general, based on the upstream and downstream measurements of proton track position and angle in case of a double-sided scanner design. Formulae are available to calculate the MLP of the protons  in~\citep{MLP_Schneider, MLPChargedParticle, MLP_Schulte, extendedMLPformalism}.

In case of a single-sided scanner design, where upstream measurements are not available, the beam information can be used. Certainly, this contains more uncertainty than a precise measurement, which must be included in the calculations---however, such a setup is more feasible with significantly lower instrumentation costs. The formalism of~\cite{MLP_Schulte} was extended by~\cite{extendedMLPformalism} to deal with the uncertainty of the measurements and the beam as well, so their formulae were applied in this work.

In order to incorporate the increased uncertainty that is originating from the missing upstream measurement and the scattering of the proton inside the patient, we have implemented a simplified probability density based approach. This was suggested by~\cite{MLPChargedParticle} and applied earlier for a double-sided setup by~\cite{Wang2010}.

The determination of the proton-phantom interaction was divided into the MLP calculation and the approximation of the probability density around the MLP. The calculation was organized into the following steps:
\begin{enumerate}[label=(\roman*)]
    \item The outgoing direction on the phantom hull is determined by the measured direction of the proton, neglecting any interactions between the first detector layer and the phantom hull. Similarly, the entry directions are determined by the known beam parameters.
    \item The perpendicular entry and outgoing coordinates are calculated based on the MLP formulae of \cite{extendedMLPformalism, MLP_Schulte} at the directional incoming and outgoing positions, respectively.
    \item From computational point of view, the MLP calculation in the whole phantom would not be feasible, therefore inside the phantom a third order spline approximation is applied, which is found to be accurate and was suggested by~\cite{MLPChargedParticle}.
    \item The probability density around the MLP is approximated by a Gaussian~\citep{MLPChargedParticle}. To simplify the calculations, the standard deviation ($\sigma$) of this Gaussian was considered constant 
    along the proton path in the phantom. This approach does not deal with the uncertainty characteristics along the proton path, but it includes the uncertainty that is originating from the scattering of the proton in the phantom.
\end{enumerate}
The proposed approach strikes a balance between physical accuracy and computational efficiency in determining the most likely paths of the protons. It offers a well-founded, physically motivated calculation, while maintaining the computational efficiency necessary for practical application at the current level, which aims to provide a proof-of-concept study of the proposed Richardson\,--\,Lucy method.

\section{Simulations with the Algorithm}
\label{sec:test}

The ultimate goal of the pCT imaging for proton therapy is to provide a solid basis for accurate and reliable dose planning. It is traditionally a challenge to define a proper measure, which correctly characterizes the goodness of a reconstructed RSP distribution, therefore general image properties (spatial and density resolution) are usually used to quantify image quality. In order to evaluate our proposed method, the imaging of dedicated spatial and density resolution phantoms were simulated with Monte Carlo techniques, then reconstructed by the formerly described method and evaluated following the instructions later on this section.

\subsection{The Proton CT Scanner Model} 
\label{subsec:pctmodel}

\begin{figure}[h]
    \centering
    \includegraphics[width=0.48\textwidth]{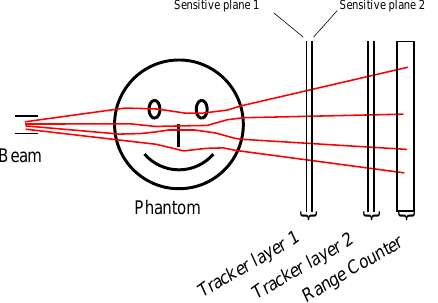}
    \caption{Single-sided, list mode detector design.}
    \label{fig:singleSidedSetup}
\end{figure}

A single-sided detector design (Figure~\ref{fig:singleSidedSetup}) with a 230 MeV/$u$ pencil beam was investigated. Following the realistic beam model of~\cite{noFrontTrack_Rambo}, the full width at half maximum (FWHM) of the Gaussian beam was set to 7~mm (with about 3 mm standard deviation), the spot divergence was chosen to be 2.8~mrad and the spot emittance was 3.0~mrad$\times$mm. 

Three different detector layer setups were compared in this study: the first is an {\sl idealized detector} with no measurement errors, the second is a {\sl silicon pixel tracker} modeled after the design of the Bergen pCT Collaboration~\citep{DesignPixelRangeTelescope, BergenpCTStatusReport}. Finally, the third is a {\sl silicon strip detector} based tracker layer, followed the LLU/UCSC Phase-II Scanner design of the Loma Linda University (LLU) and the University of California at Santa Cruz (UCSC)~\citep{LL-Phase2}.
The relevant properties of the three detector layer setups, namely the idealized, the silicon pixel, and the silicon strip, are summarized in Table~\ref{tab:layers}. 
\begin{table}[h]
\footnotesize
  \begin{center}
    \begin{tabular}{lrrr}
      \hline
      & \textbf{Ideal} & \textbf{Silicon pixel} &  \textbf{Silicon strip}\\
      \hline
      \hline
        Layer material budget ($x/X_0$) & 0 & $4.2\times10^{-3}$ & $8.5\times10^{-3}$ \\
        Distance between layers (mm) & - & 50 & 50 \\
        Spatial resolution ($\mu$m) & 0 & 5 & 66 \\
        Angular resolution (130-230 MeV/u, mrad) & 0 & 1.7-2.9 & 3.1-4.6 \\
        Correlation (130-230 MeV/u, (mrad$\times$mm) & 0 & $-5\times10^{-4}$ & $-8.7\times10^{-2}$ \\
        Statistical WEPL resolution (mm) & 0 & 3.0 & 3.0 \\
      \hline
    \end{tabular}
  \end{center}
  \caption{Comparison of tracker detector pair model parameters: Ideal setup, with no measurement errors, Silicon pixel detector based on the design of the Bergen pCT Collaboration~\citep{DesignPixelRangeTelescope, BergenpCTStatusReport}, and Silicon strip detector model following the structure of the LLU/UCSC Phase-II Scanner~\citep{LL-Phase2}. }
  \label{tab:layers}
\end{table}

\begin{figure}[h!]
    \centering
    \includegraphics[width=\linewidth]{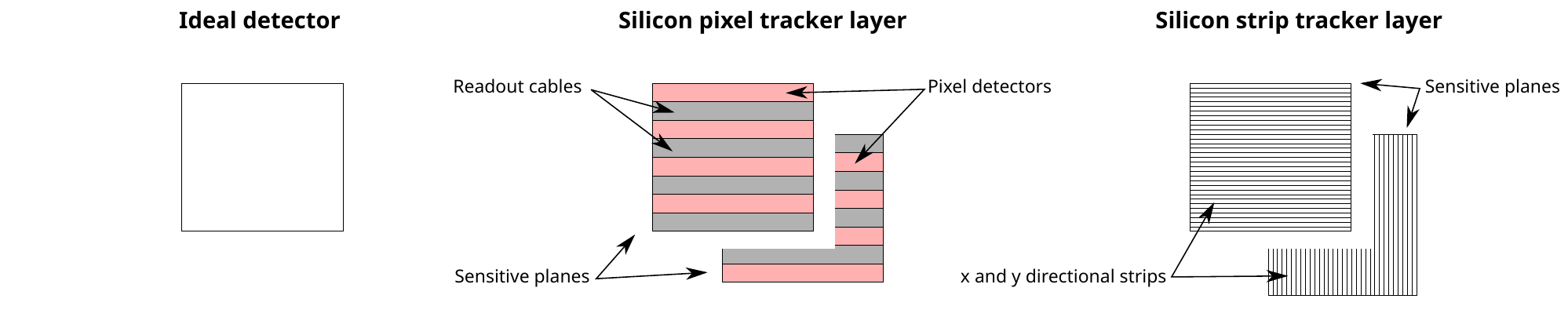}
    \caption{The structure of the investigated layers.}
    \label{fig:layers}
\end{figure}

As illustrated in Figure~\ref{fig:singleSidedSetup}, the idealized setup is a single sensitive plane, but in the latter two realistic cases each tracker layer contains two sensitive planes similarly as in the existing technological solutions. If the detection is based on a silicon pixel detector, a double structure of two equivalent sensitive planes is needed to be applied to fully cover the alternating sensitive and readout electronics panels. While applying silicon strip detectors, two separate planes are required for the perpendicular geometrical $x$ and $y$ directions. The schematic structures of the detector layers are shown in Figure~\ref{fig:layers}. Table~\ref{tab:layers} contains the joint material budget of these double layers. 

The WEPL resolution of both realistic setups was chosen to be 3 water equal mm (standard deviation of a normal distribution), which was added to the simulated range straggling in the phantom. This is a realistic uncertainty---however, this is a very rudimentary model, as the measurement error is likely to depend on the remaining range of the protons behind the patient. The distance of the first detector pair to the rotation axis (isocenter) was chosen for 400~mm in all cases, which results in 325~mm detector-phantom distance for the phantoms. Similar distances were used for portal detectors applied in photon therapy gantries, so they can be considered to be realistic for the future pCT devices as well~\citep{extendedMLPformalism}.

\subsection{The Investigated Phantoms} 
\label{subsec:phantoms}

In order to quantify the effectiveness of the proposed Richardson\,--\,Lucy algorithm, a standardized evaluation method is required. In our study, two widely-applied phantoms have been used to test the goodness of the reconstruction. The RSP distribution was reconstructed in one plane of the phantoms, so the phantoms were considered to be offset invariant in the direction of the rotation axis. To ensure the offset invariance, 400~mm thick phantoms were simulated in the axial direction.

The spatial resolution of the reconstruction was measured with the Monte Carlo imaging of the CTP528 phantom~\citep{CTP404}, which includes a high-resolution test gauge ranging from 1 to 21 line pairs per centimeter embedded within a uniform material. By minimizing the presence of high-contrast materials and utilizing a radial design, the CTP528 module effectively reduces streaking artifacts. The inserts are arranged in a circular configuration, specifically designed to facilitate the evaluation of spatial resolution through Modulation Transfer Function (MTF) analysis.

The RSP reconstruction accuracy (also referred to as density resolution) has been evaluated with the CTP404 phantom~\citep{CTP404}. It is designed to measure how accurately a material property is reconstructed in a homogeneous region of the phantom. The CTP404 phantom is a 150~mm diameter epoxy cylinder, which contains 8 different material inserts with a diameter of 12.2~mm. 

\subsection{The Workflow}
\label{subsec:workflow}

A simulation code was developed to use and test the image reconstruction algorithm, which was divided into the following steps (illustrated in Figure \ref{fig:blockDiagram}):
\begin{figure}[h]
\centering
    \includegraphics[width=0.60\linewidth]{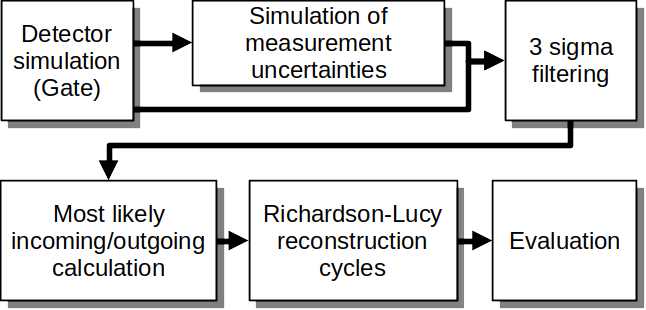}
    \caption{Simulation steps of the Richardson\,--\,Lucy algorithm evaluation.}
    \label{fig:blockDiagram}
\end{figure}
\begin{enumerate}[label=(\roman*)]

\item The data taking was simulated with Monte Carlo method. Beside the pencil beam model, the phantoms were modeled appropriately in the simulations. We used the Geant4 (version 11.0.0)~\citep{Agostinelli2003, Allison2006} with GATE (version 9.2)~\citep{Jan2004, Jan2011} to simulate the phantom-beam interactions. In the reference physics list settings QGSP\_BIC\_EMY was activated for these calculations with an ionization potential of water manually set to 75 eV. Data taking of one slice was simulated from 360 directions in $1^\circ$ and 1~mm steps, with 1000 primary proton at each position. 
In total, $\sim 43$~million primary protons were used for the reconstruction of the phantoms that didn't suffer large angle scattering. We note that this number of primary protons is compatible with a traditionally assumed low-dose imaging configuration, corresponding to roughly 0.05 mGy deposited dose. This is approximately 50 times lower than in a conventional x-ray CT~\citep{ProtonVDA, noFrontTrack_Rambo, LL-Phase2, Plautz2016}. However, these quantities may vary significantly depending on the specifics of the given imaging setup, and a more precise dose-dependent evaluation of the algorithm is out of the scope of the current study.

\item The position and direction uncertainties for the realistic detector setups were simulated using correlated Gaussian distributions, and added to the exact positions and directions of simulated protons. The measurement uncertainty was calculated based on the guideline of~\cite{extendedMLPformalism}. The WEPL measurement error also was randomly assigned from a Gaussian distribution to the WEPL of the protons calculated from their energy losses, simulated in the previous step, according to the parameters listed in table~\ref{tab:layers}.

\item In order to filter out the protons which undergo nuclear collisions in the phantom, a 3-sigma filtering has been applied for the direction and WEPL of the protons originating from the same beam spot, as it was suggested and used by~\cite{MLP_Schulte}.

\item Calculation of the most probable incoming and outgoing positions and directions (MLP) of the protons on a cylinder around the phantom hull was performed, based on the formulation of~\cite{extendedMLPformalism, MLP_Schulte}. 

\item The Richardson\,--\,Lucy-based method was used to reconstruct the RSP distribution from the individual proton histories. On-the-fly system matrix calculation was applied based on simplified Gaussian probability density around a third order spline approximation of the MLP, where the probability density is characterized by the $\sigma$ variance of the Gaussian (chosen in the order of the voxel's size).

\item In the final step, the spatial resolution was evaluated based on the reconstruction of the CTP528 phantom. The RSP accuracy was calculated from the reconstructed CTP404 phantom.
\end{enumerate}

Calculations were done on the machines of the Wigner Scientific Computing Laboratory's hardware. The computationally demanding part of the algorithm was running on Nvidia 1080~Ti Graphical Processing Unit (GPU) cards. The algorithm has been designed so that the list of all available proton tracks are grouped and processed in small chunks (typically consisting $2\times 10^4$ protons) that fit in the memory of the used GPU device. Within a chunk, the weights of the iterations defined by Eq.~(\ref{eq:RL_pct}) should be calculated once and only once~\citep{Erdogan_1999}. The iteration on each chunk is processed until a given number of $k$ cycles is reached, or a certain, adaptively set stopping condition is achieved (typically after 20-30 cycles). In the current study, this condition is simply defined as a threshold in the mean squared error (MSE) between two consecutive cycles. 
This ensures that the processing of the given chunk is terminated and the signal-to-noise ratio kept maximal, while the computation time is optimal. In a future development further, more robust metrics can be considered as well, such as the structural similarity index measure (SSIM) and peak signal-to-noise ratio (PSNR)---in the current study we have found that the described method with the MSE already leads to satisfactory results.

A well-known limitation of the traditional Richardson–Lucy–based iterative schemes is that increasing the number of iterations tends to amplify high-frequency noise ~\citep{rl_noise}. While this effect can be mitigated by the aforementioned appropriate stopping criteria or regularization strategies, an additional stabilization mechanism is employed in the present workflow. Specifically, the output of each batch is used as the input to the subsequent batch, but the intermediate reconstructions are also accumulated and averaged. This ensemble averaging of partial results significantly improves the signal-to-noise ratio while largely preserving the resolution gains achieved by the RL iterations.

Finally, the contour of the (generally unknown) convex hull of the phantom can be determined more precisely with the increasing number of processed tracks, improving the contrast around the edges.

\subsection{Evaluation of the CTP528 Phantom}
\label{subsec:Eval_Derenzo}

Spatial resolution is a critical parameter in evaluating the performance of an image reconstruction algorithm. It refers to the ability of the imaging system to distinguish between small structures in the object being imaged. In essence, it measures the contrast between adjacent features---higher contrast implies that smaller structures can be differentiated effectively. Typically, spatial resolution is quantified using metrics such as line pairs per millimeter (lp/mm) or line pairs per centimeter (lp/cm), where a higher number of line pairs indicates a better resolution~\citep{warren2007, nadia2011}.

In any imaging system, the inherent effect is similar to a blurring of the image, where fine details become smeared out. This blurring can be mathematically characterized by the Point Spread Function (PSF), which describes how a single point in the object space is distributed in the image space. The PSF essentially encapsulates the extent of the blurring effect introduced by the imaging system.

To analyze the frequency behavior of the system, the Fourier transform of the PSF is computed, resulting in the Modulation Transfer Function (MTF). The MTF provides a comprehensive description of how different spatial frequencies (i.e., the rate of change of intensity in the image) are transferred by the imaging system. High spatial frequencies correspond to finer details in the image, and a higher MTF value at these frequencies indicates that the system can accurately reproduce those details.

The spatial resolution is commonly defined by the point on the MTF curve where the MTF value drops to 10\% of its maximum. This MTF(10\%) value represents the spatial frequency beyond which the imaging system's ability to differentiate between structures diminishes significantly, serving as a standard benchmark for comparing the resolution of different imaging systems or reconstruction algorithms. This method to acquire the MTF(10\%) values was motivated by~\cite{noFrontTrack_Rambo}, however we note that various evaluation methods may provide slightly different spatial resolution values.

For the evaluation of the reconstructed image of the CTP528 phantom, an automatized strategy has been developed based on~\cite{noFrontTrack_Rambo} and~\cite{piersimoni2018helium}. The main steps of the evaluation follows as:

\begin{enumerate}[label=(\roman*)]
    \item The reconstructed RSP matrix is rotated such that the Region of Interest (ROI) consistently appears in the same position. The rotation angles of the inserts are specified in the GATE macro of the phantom.  
    \item The ROI is extracted, and line profiles are recorded along several projections to reduce uncertainties (see Figure \ref{fig:ctp528_workflow}).  
\begin{figure}[h]
    \centering
    \includegraphics[width=0.6\textwidth]{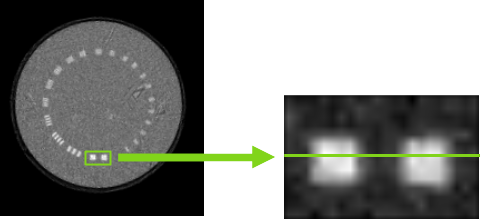}
    \caption{Extracting the RSP profile along a line of the Region of Interest at the 1~lp/cm region of CTP528 phantom.}
    \label{fig:ctp528_workflow}
\end{figure}
    \item The local minimum and maximum values of the line profiles are determined and subsequently averaged (see Figure \ref{fig:ctp528_extrema} as an illustration).  
\begin{figure}[h]
    \centering
    \begin{subfigure}{.49\textwidth}
        \includegraphics[width=\textwidth]{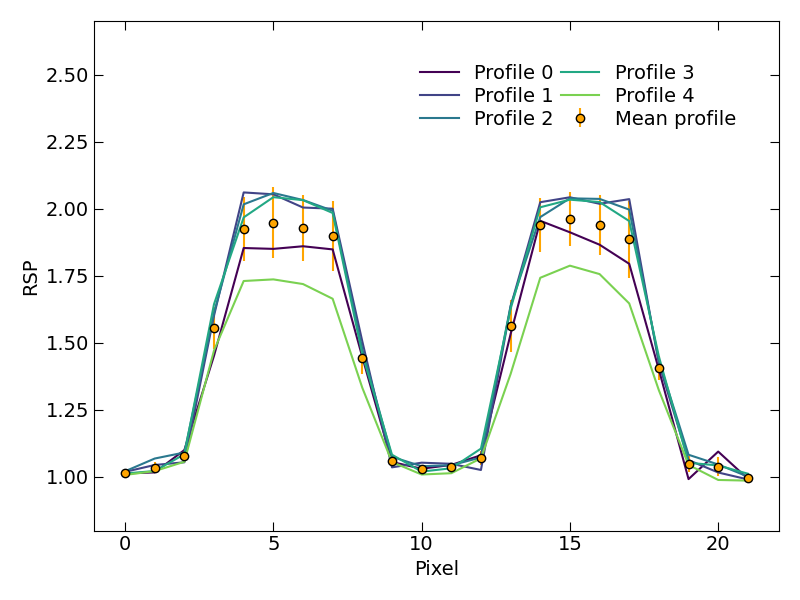}
        \caption{}
        \label{subfig:ctp528_profiles}
    \end{subfigure}
    \begin{subfigure}{.49\textwidth}
        \includegraphics[width=\textwidth]{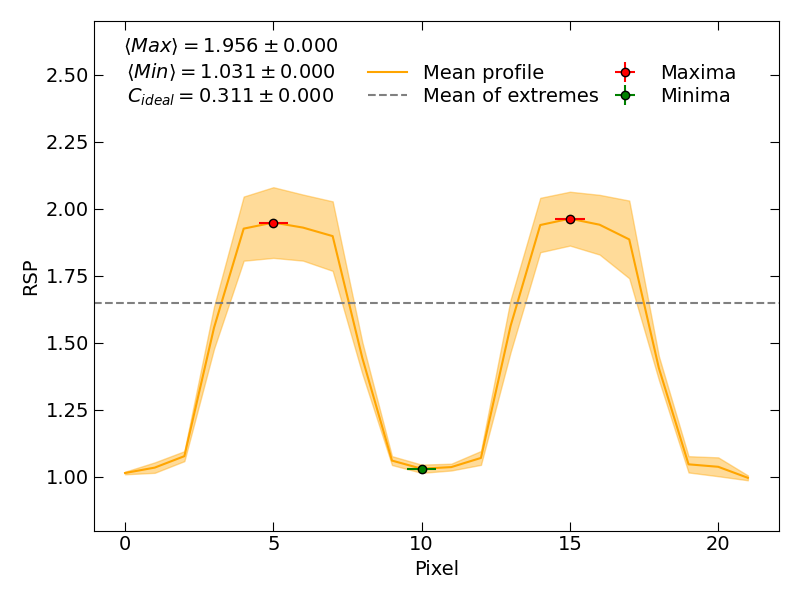}
        \caption{}
        \label{subfig:ctp528_smoothed}
    \end{subfigure}
    \caption{The intensity profiles and the found minima/maxima for the 1 lp/cm segment of the reconstructed CTP528 phantom.}
    \label{fig:ctp528_extrema}
\end{figure}
    \item The contrast is computed from all minimum and maximum pairings as:  
\begin{equation}
    C(f)=\biggl<\frac{RSP_{max}-RSP_{min}}{RSP_{max}+RSP_{min}}\biggr>.
\end{equation}
    \item The MTF is calculated for all investigated insert sizes as the following:
\begin{equation}
    MTF(f)=\frac{C(f)}{C(0)},
\end{equation}
where $C(0)$ is taken as the contrast between the ground truth RSP values of aluminum and water.
    \item Finally, a parametrized sinc function is fitted to the contrast values in order to mitigate the uncertainty, and the 10\% threshold is extracted from the resulting fit as the spatial resolution of the image reconstruction system.  
\end{enumerate}

\subsection{Evaluation of the CTP404 Phantom}
\label{subsec:Eval_CTP404}

\begin{figure}[h]
    \centering
    \begin{subfigure}{.40\textwidth}
        \includegraphics[width=\textwidth]{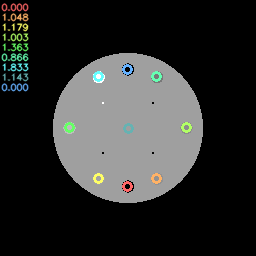}
        \caption{}
        \label{subfig:ctp404_contoured_gt}
    \end{subfigure}
    \begin{subfigure}{.40\textwidth}
        \includegraphics[width=\textwidth]{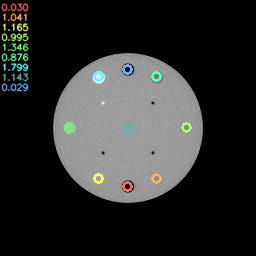}
        \caption{}
        \label{subfig:ctp404_contoured_ideal}
    \end{subfigure}
    \caption{The ground truth and ideal reconstructions of the CTP404 phantom with the area averaged RSP regions.}
    \label{fig:ctp404_contoured}
\end{figure}

In order to robustly determine the average RSP of the inserts on the reconstructed images, yet an another automatized algorithm has been implemented as the following:
\begin{enumerate}[label=(\roman*)]
    \item At given reconstruction resolution (determined by the mm/pixel values), the exact center position of each insert is determined on the ground truth image (which has perfect contrast, zero noise and blurring).
    \item On the reconstructed CTP404 images, the mean RSP around the previously determined center positions was calculated---see Figure \ref{fig:ctp404_contoured} as an illustration for the ground truth and ideal reconstructions, where the colored sections indicate the reconstructed RSP values of the inserts.
\end{enumerate}
The extracted RSP values can be compared to the ground truth RSP values, as it was presented in details by~\cite{BergenpCTStatusReport}. 

\section{Results \label{sec:results}} 

In this study the proof-of-concept performance of the Richardson\,--\,Lucy-based reconstruction algorithm was investigated using a single-sided detector setup, simulated by a simplified Monte Carlo model. The center line of all beams coincided in one plane, perpendicular to the rotation axis. Moreover, every proton was assigned into this layer, without taking into account the deviation of their path in the direction of the rotation axis. This 256$\times$256 image slice was reconstructed using 1~mm/pixel reconstruction resolution. For this resolution, the variance of the probability density around the MLP was chosen to be $\sigma=0.4$~mm.
\begin{figure}[h]
    \centering
    \includegraphics[width=0.32\textwidth]{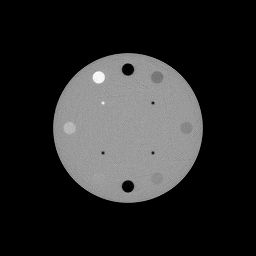}
    \includegraphics[width=0.32\textwidth]{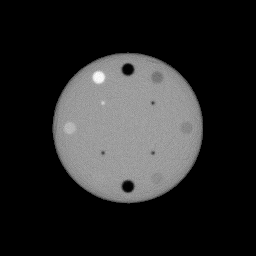}
    \includegraphics[width=0.32\textwidth]{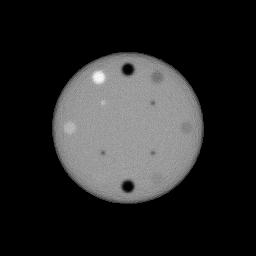}
    \includegraphics[width=0.32\textwidth]{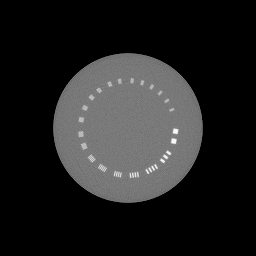}
    \includegraphics[width=0.32\textwidth]{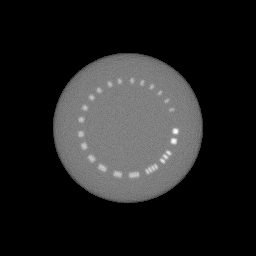}
    \includegraphics[width=0.32\textwidth]{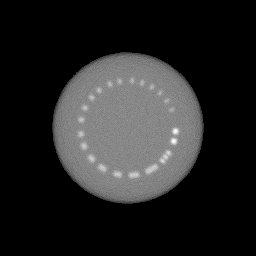}
    \caption{Richardson\,--\,Lucy algorithm based reconstruction of the CTP404 (top row) and CTP528 (bottom row) phantoms.  The ideal, the pixel detector, and the strip detector layers are shown, respectively from left to right.}
    \label{fig:recons}
\end{figure}

The result of the reconstructed images of the CTP404 and CTP528 phantoms are shown in Figure \ref{fig:recons}, on the top and bottom rows respectively. 
The left-hand side column of Figure~\ref{fig:recons} presents the idealized case, while the middle and right-hand side columns show the more realistic silicon pixel and silicon strip detector models, respectively. 

\begin{figure}[h]
    \centering
    \begin{subfigure}{.49\textwidth}
        \includegraphics[width=\textwidth]{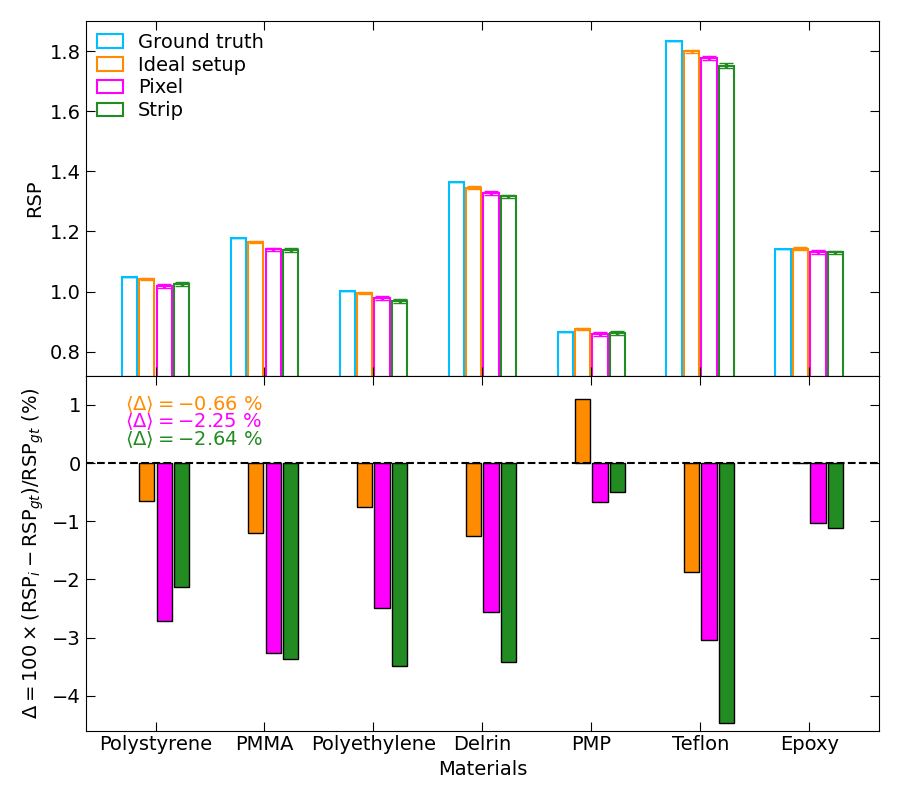}
        \caption{}
        \label{subfig:rsp}
    \end{subfigure}
    \begin{subfigure}{.49\textwidth}
        \includegraphics[width=\textwidth]{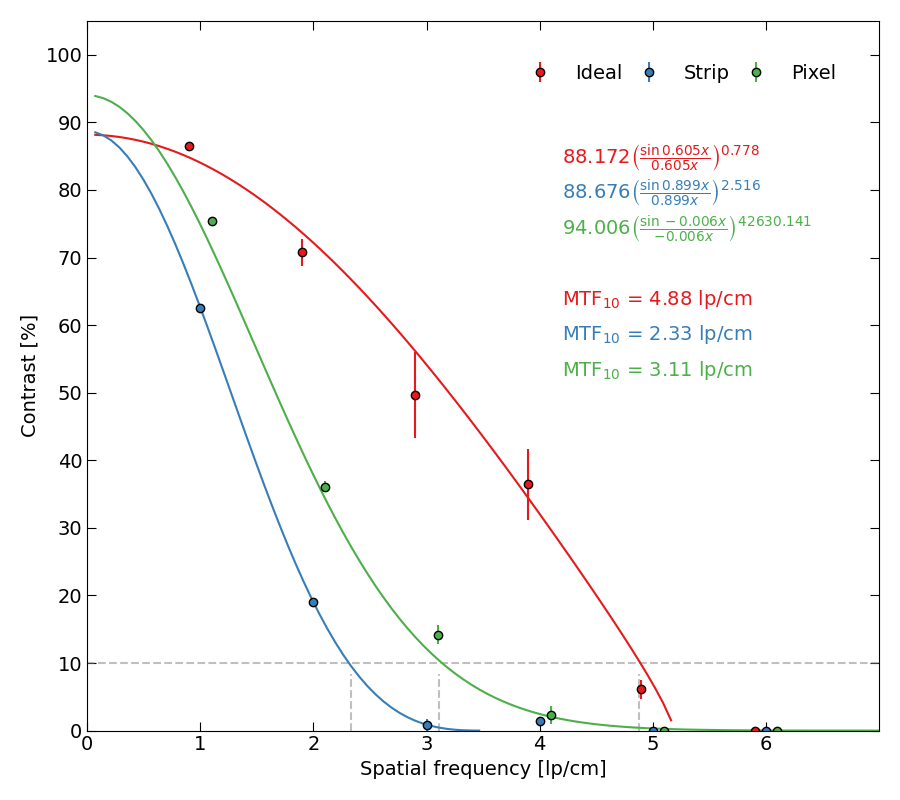}
        \caption{}
        \label{subfig:mtf}
    \end{subfigure}
    \caption{Left panel: The reconstructed RSP values from the CTP404 phantom and their relative differences compared to the ground truth values, for the various material inserts. Right panel: the average spatial resolution in the function of the processed proton tracks.}
    \label{fig:comparisons}
\end{figure}

To quantify the quality of the reconstructed images, the RSP accuracy and its relative difference is shown in Figure~\ref{subfig:rsp}, while Figure~\ref{subfig:mtf} shows the spatial resolution (as defined in Subsection~\ref{subsec:Eval_Derenzo}). The reconstructed RSP values and their errors are summarized in Table~\ref{tab:RSPdifferences}.

The average relative RSP difference for the listed materials was found to be -0.66\% for the ideal, -2.25\% for the silicon pixel, and -2.64\% for the silicon strip setups after processing 43 million protons.
This average accuracy exceeds the required 1\% uncertainty for the ideal setup~\citep{accuracy}.

After processing 43 million protons, the spatial resolution was found to be 4.88~lp/cm for the ideal setup, 3.11~lp/cm and 2.33~lp/cm for the silicon pixel and the silicon strip detector based setups, respectively. As it is visible also on the middle and right panels of Figure~\ref{fig:recons}, the detector uncertainties result in a more blurred final image.
\begin{table}[h]
\scriptsize
  \begin{center}
    \begin{tabular}{lrrrrrrrrrrrrr}
      \hline
\textbf{Insert} & \multirow{2}{*}{\shortstack{\textbf{RSP}\\(gr. tr.)}}   & \multicolumn{2}{l}{\textbf{Ideal setup}}  & & \multicolumn{2}{l}{\textbf{Silicon pixel}}   & &  \multicolumn{2}{l}{\textbf{Silicon strip}} & \\
                            & &   \multirow{2}{*}{RSP}  & Std.    & Rel. & \multirow{2}{*}{RSP} & Std.    & Rel. & \multirow{2}{*}{RSP} & Std.   & Rel. \\
                          & &   & err. . & diff. &  & err. & diff.& & err.  & diff. \\
      \hline
      \hline
Polystyrene  &	1.048 &	1.041 &	0.004 &		-0.007 &	1.020 &	0.006  &	-0.027 &	1.026 &	0.006 &	-0.021 \\
PMMA         &	1.179 &	1.165 &	0.004 &	    -0.012 &	1.140 &	0.005  &	-0.033 &	1.139 &	0.006 &	-0.034 \\
Polyethylene &	1.003 &	0.995 &	0.004 &		-0.008 &	0.978 &	0.006  &	-0.025 &	0.968 &	0.006 &	-0.035 \\
Delrin       &	1.363 &	1.346 &	0.005 &		-0.013 &	1.328 &	0.006  &	-0.026 &	1.316 &	0.006 &	-0.034 \\
PMP          &	0.866 &	0.876 &	0.005 &		 0.011 &	0.860 &	0.007  &	-0.007 &	0.862 &	0.007 &	-0.005 \\
Teflon       &	1.833 &	1.799 &	0.004 &		-0.019 &	1.777 &	0.006  &	-0.030 &	1.751 &	0.008 &	-0.045 \\
Epoxy        &	1.143 &	1.143 &	0.004 &		0.000  &	1.131 &	0.006  &	-0.010 &	1.130 &	0.006 &	-0.011 \\
Air-2        &	0.001 &	0.029 &	0.002 &		-      &	0.026 &	0.002  &	-      &	0.039 &	0.004 &	- \\
Air-1        &	0.001 &	0.030 &	0.002 &		-      &	0.025 &	0.002  &	-      &	0.038 &	0.004 &	- \\
      \hline
    \end{tabular}
  \end{center}
  \caption{The difference between the real and the reconstructed RSP of the eight inserts. The epoxy data was measured in a 4~mm radius circle in the middle of the phantom.}
  \label{tab:RSPdifferences}
\end{table}

\section{Discussion and Comparison} 
\label{sec:discussion}

Our reconstruction for the CTP528 phantom shows that 4.88~lp/cm, 3.11~lp/cm and 2.33~lp/cm resolution can be achieved for the ideal, pixel and strip detector models respectively. With this, we were able to achieve the 97\% of the limiting Nyquist frequency for the 1~mm/px resolution for the ideal case, while 62\% and 47\% for the pixel and strip detector models respectively. 

In \cite{noFrontTrack_Rambo} the authors have reached a 3.8~lp/cm spatial resolution with the ideal and 3.2~lp/cm spatial resolution with a realistic single-sided detector setup, whereas they have applied a higher pixel density of 0.25~mm/px at the reconstruction. 
In~\cite{Collins-Fekete_2016}, the highest reported value for spatial resolution was 5.76~lp/cm for 330 MeV beams, without detector effects, using 10 million primary protons per rotation angle.

These results are already promising and reaching the minimally viable values for treatment planning~\citep{extendedMLPformalism}. The current, proof-of-concept state method can be extended to handle further effects (such as variable beam energies, different most likely paths calculation methods, etc) and they are planned for a future study. Note that our achieved results should be compared directly to single-sided setups only---naturally, a double-sided detector system could achieve a better resolution up to 40\% \citep{noFrontTrack_Rambo}, along with the drawback of significantly higher instrumental costs.

In \cite{ProtonVDA} the authors have achieved a robust RSP accuracy using a double-sided detector setup and 3 dimensional measurements, with uncertainties up to 1-2\% in soft tissues and 7\% in bone.
Our achieved RSP accuracy was already found to be similarly close to the required 1\% RSP uncertainty with the investigated single-sided detector models, which also indicates the relevance of the proposed algorithm. However, there is room for improvement in this regard as well: for example, the currently applied simplified probability density model could be superseded with a more detailed probability map~\citep{Wang2010}, while a true 3-dimensional reconstruction is required to further improve the result. 

\begin{figure}[h]
    \centering
    \begin{subfigure}{.35\textwidth}
        \includegraphics[width=\textwidth]{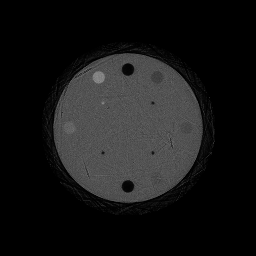}
        \caption{}
        \label{subfig:ctp404_partial}
    \end{subfigure}
    \begin{subfigure}{.54\textwidth}
        \includegraphics[width=\textwidth]{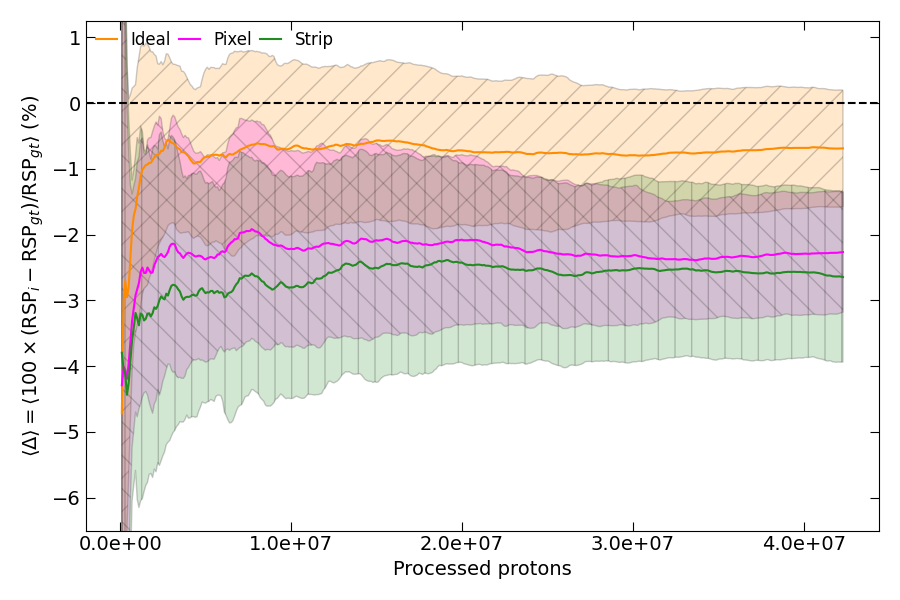}
        \caption{}
        \label{subfig:ctp404_delta}
    \end{subfigure}
    \caption{Left panel: The reconstructed CTP404 phantom for the ideal setup after 4 million proton tracks. Right panel: the mean relative difference of the calculated RSP in the function of the processed proton tracks.}
    \label{fig:ctp404_partial}
\end{figure}

The obtained resolution was reached by processing a high number of the primary protons with a fine spatial coverage, calculated within the order of 300 minutes running on a common, commercial GPU card (the overhead of the approximated MLP calculation is negligible due to the straightforward CPU parallelization). Although, it is important to remark that for both phantoms a visually satisfying result was achieved already after processing only $\sim$10\% of the full data (i.e. 4 million proton tracks), with 20-22 minutes of computational time. Figure \ref{fig:ctp404_partial} illustrates the partial reconstruction with the mean of the relative RSP differences with respect to the processed proton numbers. Certainly, using a stronger GPU could immediately lead us to increase the batch number of protons, and therefore improved resolution and runtime. However, in the current proof-of-concept development phase only a handful of computational optimization techniques have been applied, with the main emphasis on the feasibility study of the method instead of computational efficiency. Therefore, with future developments the algorithm will provide excellent potential to become a competitive and clinically viable method.

\section{Summary}
\label{sec:summary} 

In this work the first application of the Richardson\,--\,Lucy-based iterative algorithm with probability density based proton-phantom interaction calculation for proton CT image reconstruction has been presented. We applied clinically relevant setups and parameters in the Monte Carlo simulations: 400~mm detector-isocenter distance, realistic beam and detector characteristics, with $\sim$40 million primary protons per image slice. In order to test and evaluate the reconstruction results, two widely used phantoms were applied: the CTP528 and the CTP404 ones.

We have concluded that the presented reconstruction method already approaches the required density uncertainty and spatial resolution, with similar values as the state-of-art prototypes. As of today, quantitatively more accurate or computationally more effective methods are available in e.g.~\cite{noFrontTrack_Rambo} and~\cite{Collins-Fekete_2016}---however, the proposed method has the potential to achieve qualitatively competitive results and medical applicability. 

We plan to continue the development of the proposed algorithm, with a focus on the spatial resolution, three dimensional reconstruction and improved processing time, which limited the investigations of the current work to only one layer. Indeed, the possibility of the speedup of the algorithm has been also directed by parallel booster methods and machine learning techniques.

\section{Declaration of generative AI and AI-assisted technologies in the writing process}
During the preparation of this work the authors used ChatGPT 4o in order to rephrase certain sentences for improved readability. After using
this tool/service, the author reviewed and edited the content as needed and take full responsibility for the content of the published article.

\section*{Acknowledgement}

This work has been supported by the Hungarian National Research, Development and Innovation Office (NKFIH) under the contract numbers {NKFIH} NKKP ADVANCED\_25-153456, NEMZ\_KI-2022-00058, 2025-1.1.5-NEMZ\_KI-2025-00005, 2025-1.1.5-NEMZ\_KI-2025-00013 and 2024-1.2.5-TET-2024-00022, the FuSe COST Action CA-24101 and the Wigner Scientific Computing Laboratory (WSCLAB). Author Zs. J. has been supported by the EKÖP-25 University Research Scholarship Program of the Ministry for Culture and Innovation from the source of the National Research, Development and Innovation Fund. This work was also supported by the Research Council of Norway (Norges forskningsrad) and the University of Bergen, grant number 250858. The authors acknowledge the support of Trond Mohn Foundation (BFS2017TMT07).

\section*{Members of the Bergen pCT Collaboration}

\footnotesize

M. Aehle\textsuperscript{a}, 
J. Alme\textsuperscript{b}, 
G.G. Barnaföldi\textsuperscript{c},
G. Bíró\textsuperscript{c,l},
T. Bodova\textsuperscript{b}, 
V. Borshchov\textsuperscript{d}, 
A. van den Brink\textsuperscript{b}, 
M. Chaar\textsuperscript{b}, 
B. Dudás\textsuperscript{l}, 
V. Eikeland\textsuperscript{e}, 
G. Feofilov\textsuperscript{f}, 
C. Garth\textsuperscript{g}, 
N.R. Gauger\textsuperscript{a}, 
O. Grøttvik\textsuperscript{b}, 
H. Helstrup\textsuperscript{h}, 
S. Igolkin\textsuperscript{f}, 
Zs. Jólesz\textsuperscript{c,l},
R. Keidel\textsuperscript{i}, 
C. Kobdaj\textsuperscript{j},
T. Kortus\textsuperscript{a}, 
L. Kusch\textsuperscript{v}, 
V. Leonhardt\textsuperscript{g}, 
S. Mehendale\textsuperscript{b}, 
R. Ningappa Mulawade\textsuperscript{i}, 
O.H. Odland\textsuperscript{k, b}, 
G. O'Neill\textsuperscript{b}, 
G. Papp\textsuperscript{l}, 
T. Peitzmann\textsuperscript{e}, 
H.E.S. Pettersen\textsuperscript{k}, 
P. Piersimoni\textsuperscript{b,m}, 
M. Protsenko\textsuperscript{d}, 
M. Rauch\textsuperscript{b}, 
A. Ur Rehman\textsuperscript{b}, 
M. Richter\textsuperscript{n}, 
D. Röhrich\textsuperscript{b}, 
J. Santana\textsuperscript{i}, 
A. Schilling\textsuperscript{a}, 
J. Seco\textsuperscript{o, p}, 
A. Songmoolnak\textsuperscript{b, j}, 
Á. Sudár\textsuperscript{x, y},
J. Rambo Sølie\textsuperscript{q}, 
G. Tambave\textsuperscript{w}, 
I. Tymchuk\textsuperscript{d}, 
K. Ullaland\textsuperscript{b}, 
M. Varga-Kőfaragó\textsuperscript{c}, 
L. Volz\textsuperscript{s}, 
B. Wagner\textsuperscript{b}, 
S. Wendzel\textsuperscript{i}, 
A. Wiebel\textsuperscript{i}, 
R. Xiao\textsuperscript{b, t}, 
S. Yang\textsuperscript{b}, 
H. Yokoyama\textsuperscript{e}, 
S. Zillien\textsuperscript{i}

a) Chair for Scientific Computing, University of Kaiserslautern-Landau, 67663 Kaiserslautern, Germany; \\
b) Department of Physics and Technology, University of Bergen, 5007 Bergen, Norway; \\
c) HUN-REN Wigner Research Centre for Physics, 29--33 Konkoly--Thege Mikl\'os \'ut,\\ H-1121 Budapest, Hungary; \\
d) Research and Production Enterprise "LTU" (RPELTU), Kharkiv, Ukraine; \\
e) Institute for Subatomic Physics, Utrecht University/Nikhef, Utrecht, Netherlands; \\
f) St. Petersburg University, St. Petersburg, Russia; \\
g) Scientific Visualization Lab, University of Kaiserslautern-Landau, 67663 Kaiserslautern, Germany; \\
h) Department of Computer Science, Electrical Engineering and Mathematical Sciences, Western Norway University of Applied Sciences, 5020 Bergen, Norway; \\
i) Center for Technology and Transfer (ZTT), University of Applied Sciences Worms, Worms, Germany; \\
j) Institute of Science, Suranaree University of Technology, Nakhon Ratchasima, Thailand; \\
k) Department of Oncology and Medical Physics, Haukeland University Hospital, 5021 Bergen, Norway; \\
l) Institute for Physics, Eötvös Loránd University, 1/A Pázmány P. Sétány, H-1117 Budapest, Hungary; \\
m) UniCamillus - Saint Camillus International University of Health Sciences, Rome, Italy; \\
n) Department of Physics, University of Oslo, 0371 Oslo, Norway; \\
o) Department of Biomedical Physics in Radiation Oncology, DKFZ—German Cancer Research Center, Heidelberg, Germany; \\
p) Department of Physics and Astronomy, Heidelberg University, Heidelberg, Germany; \\
q) Department of Diagnostic Physics, Division of Radiology and Nuclear Medicine, Oslo University Hospital, Oslo, Norway; \\
r) Budapest University of Technology and Economics, Budapest, Hungary; \\
s) Biophysics, GSI Helmholtz Center for Heavy Ion Research GmbH, Darmstadt, Germany; \\
t) College of Mechanical \& Power Engineering, China Three Gorges University, Yichang, People's Republic of China;\\
u) Department of Radiology, Faculty of Medicine, Chulalongkorn University, 1873 Rama IV Rd, Pathum Wan, Bangkok, 10330, Thailand;\\
v) Eindhoven University of Technology, Eindhoven, Netherlands;\\
w) Center for Medical and Radiation Physics (CMRP), NISER Bhubaneswar 752050, India;\\
x) National Institute of Oncology, Ráth György utca 7-9, H-1122 Budapest, Hungary;\\
y) Budapest University of Technology and Economics, Institute of Nuclear Techniques, Műegyetem rakpart 9, H-1111 Budapest, Hungary.

\normalsize

\bibliographystyle{elsarticle-num-names} 
\bibliography{pCT_Richardson_Lucy_2024.bib}

\end{document}